\documentclass{emulateapj}

\usepackage{graphicx}
\usepackage{epstopdf}
\usepackage{epsfig}

\slugcomment{}

\shorttitle{KU Cyg 5-yr dust accretion event}
\shortauthors{Tang et al.}

\begin{document}


\title{DASCH on KU Cyg: a $\sim$5 year dust accretion event in $\sim$1900}


\author{Sumin Tang, Jonathan Grindlay,  Edward Los, and Mathieu Servillat}
\affil{Harvard-Smithsonian Center for Astrophysics, 60 Garden Street, Cambridge, MA 02138}

\email{stang@cfa.harvard.edu}


\begin{abstract}
KU Cyg is an eclipsing binary consisting of a F-type star accreting through a large accretion disk from a K5III red giant.
Here we present the discovery of a 5-yr dip around 1900 found from its 100 yr DASCH light curve.
It showed a $\sim0.5$ mag slow fading from 1899 to 1903, and brightened back around 1904 on a relatively shorter timescale.
The light curve shape of the $1899-1904$ fading-brightening event differs from the dust production and dispersion process observed in R Coronae Borealis (RCB) stars,
which usually has a faster fading and slower recovery,
and for KU Cyg is probably related to the accretion disk surrounding the F star.
The slow fading in KU Cyg is probably caused by increases in dust extinction in the disk,
and the subsequent quick brightening may be due to the evaporation of dust transported inwards through the disk.
The extinction excess which caused the fading may arise from increased mass transfer rate in the system,
or from dust clump ejections from the K giant.
\end{abstract}



\section{Introduction}

KU Cyg is an Algol-type eclipsing binary system with an orbital period, P$=38.4$ days (Popper 1964, 1965).
The gainer is a 3.85 $M_\odot$ F$-$type star with a near main-sequence surface gravity, but is over-massive for its luminosity,
and is surrounded by a large accretion disk (Olson et al. 1995).
The donor is a 0.48 $M_{\odot}$ K5III red giant which probably fills its Roche Lobe (Olson et al. 1995).
The accretion disk around the gainer led to abnormally wide secondary eclipses,
and is found to be large (close to or filling the primary star Roche lobe), thick, dusty, and probably eccentric (Olson 1988; Zola 1992; Smak \& Plavec 1997).
Double-peaked H$\alpha$ emission is visible at all orbital phases with chaotic fluctuations,
which strongly supports the presence of a disk (Olson 1991; Olson \& Etzel 1995).
The inclination of the binary orbital plane is $\sim86-86.5$ degree (Olson et al. 1995; Smak \& Plavec 1997).

Here we report the discovery of a 5-yr dip in brightness of KU Cyg
between 1899 and 1904 as determined from the Digital Access to a Sky Century at Harvard (DASCH) project.
DASCH is a project to digitize and analyze
the scientific data contained in the $\sim530,000$ Harvard College Observatory (HCO) plates
taken between the 1880s and 1990s, which is a unique resource for studying temporal
variations in the universe on $\sim10-100$ yr timescales (Grindlay et al. 2009).
We have developed the astrometry and photometry pipeline, scanned $\sim$ 13,000 plates
in several selected fields (Laycock et al. 2010; Los et al. 2011; Servillat et al. 2011; Tang et al. 2011),
and discovered new types of variable stars (e.g. Tang et al. 2010).
We present the DASCH light curve of KU Cyg in section 2. Discussion is given in section 3.

\section{DASCH light curve}
The Kepler field is one of our primary target fields for our initial scanning,
selected in order to take advantage of the unprecedented Kepler data on short timescales (Borucki et al. 2010) which complements DASCH data on long timescales.
We have scanned $\sim3000$ plates in or covering part of the Kepler field.
These plates cover 5$-$25 degrees on a side with typical limiting magnitudes $13-15$ mag,
and are mostly blue sensitive (Tang et al. 2011).
KU Cyg (RA$=$20:12:45.11, Dec$=$+47:23:41.4, J2000)
is a few degrees outside the Kepler field of view, but fortunately, is still covered by the Kepler Input Catalog (KIC; Brown et al. 2011).
The KIC ID of KU Cyg is 10311340 with $g=11.675$ and $g-r=0.622$.
We used the KIC for photometric calibration, and the measurements are calibrated to the $g$ band.
We have $827$ plates taken from 1890 to 1990 with good calibrations covering KU Cyg, which yield $800$ detections and $27$ upper limits.
The typical photometric uncertainty is $\sim0.1-0.15$ mag.
Each measurement is locally re-calibrated using $38$ clean neighbor stars with similar magnitudes and colors ($g=10.5-13$ and $0.2<g-r<1$) within 20 arcmin.
More details on DASCH photometry and calibration are described in Laycock et al. (2010) and Tang et al. (2011).

The resulting DASCH light curve of KU Cyg is shown in Figure 1.
The upper panel shows the whole light curve from 1890 to 1990.
The lower-left panel shows the light curve from 1897 to 1906 to better display the fading-brightening event.
The lower-right panel shows the light curve from 1980 to 1990 binned in 1 year increments,
where error bars represent the standard deviation of the mean,
to better illustrate its yearly variability
(primary eclipses are excluded; the secondary eclipses are broad and shallow and thus not excluded).
Black dots with errorbars are DASCH measurements, and blue arrows (colors in electronic version) are upper limits.
Red circles mark primary eclipses with orbital phase $0.965-0.035$,
using the ephemeris from Olson (1988) of $\texttt{JD} = 2433884.840+38.439484\texttt{E}$.
In all three panels, blue dashed lines show a constant magnitude of 11.53, which is the median magnitude
of KU Cyg from 1910 to 1990  with phases $0.05-0.95$ (excluding the primary eclipses).
All the measurements and upper limits fainter than 12.5 mag (i.e. 1 mag below its normal high state) were taken during the primary eclipses,
well in agreement with the ephemeris given in Olson (1988).

The most striking feature in the light curve is the 5-year dip around 1900, as shown in Figure 1.
KU Cyg started to show fading around 1899, and dropped from $\sim11.4$ mag (as before 1895)
to $\sim12.0$ mag in $1902-1903$, and then in about half a year, it brightened to $\sim11.5$ mag in early 1903.
After that, it showed fluctuations between $11.5-12.0$ mag from 1903 to 1905.

KU Cyg is also variable on year-timescales outside primary eclipses with fluctuations $\sim0.1$ mag,
as shown in the yearly binned light curve in the lower-right panel of Figure 1.
Note there is a $\sim0.15$ mag brightening bump around 1984, which has been discovered by Olson (1988), and is verified by DASCH.
There seems to be a slight trend of $0.1-0.2$ mag brightening from 1910 to 1990,
however, given our systematic uncertainty over 100 years of $\sim0.1$ mag (Tang et al. 2011), it is not convincing.
Compared with its $\sim 2000$ neighbors with $g=11-12$ within 5 degrees,
the brightening trend from 1910 to 1990 in KU Cyg is only at the $2\sigma$ level.
We have also visually checked the light curves of all these neighbors,
and none of them showed similar dip around 1900 as KU Cyg, which further supports that the 5-year dip is real.

The light curve of KU Cyg, folded on the 38.439484 days period (Olson 1988) is shown in Figure 2, with the upper panel for $1910-1990$ and the lower panel for $1890-1910$.
Black dots are DASCH measurements, and blue arrows are upper limits.
Both the deep primary eclipse and the shallow, wide secondary eclipse can be seen in the folded light curve from 1910 to 1990.
KU Cyg showed larger variability on non-eclipse phases ($0.05-0.95$) during $1890-1910$ than during $1910-1990$,
with rms of 0.20 mag and 0.15 mag, respectively.
The 0.15 mag rms on non-eclipse phases during $1910-1990$ further illustrates that our typical photometric errors are below $0.15$ mag.
Note that our coverage during eclipse phases (Fig. 2) would allow a period measurement precision of $dP \sim P^2 dN/T \sim 4 \times 10^{-4}$ days for a phase uncertainty $dN \sim 0.01$ over the $T\sim100$ years duration and so it is not possible to further refine the binary period given by Olson (1988).

\begin{figure*}
\epsfig{file=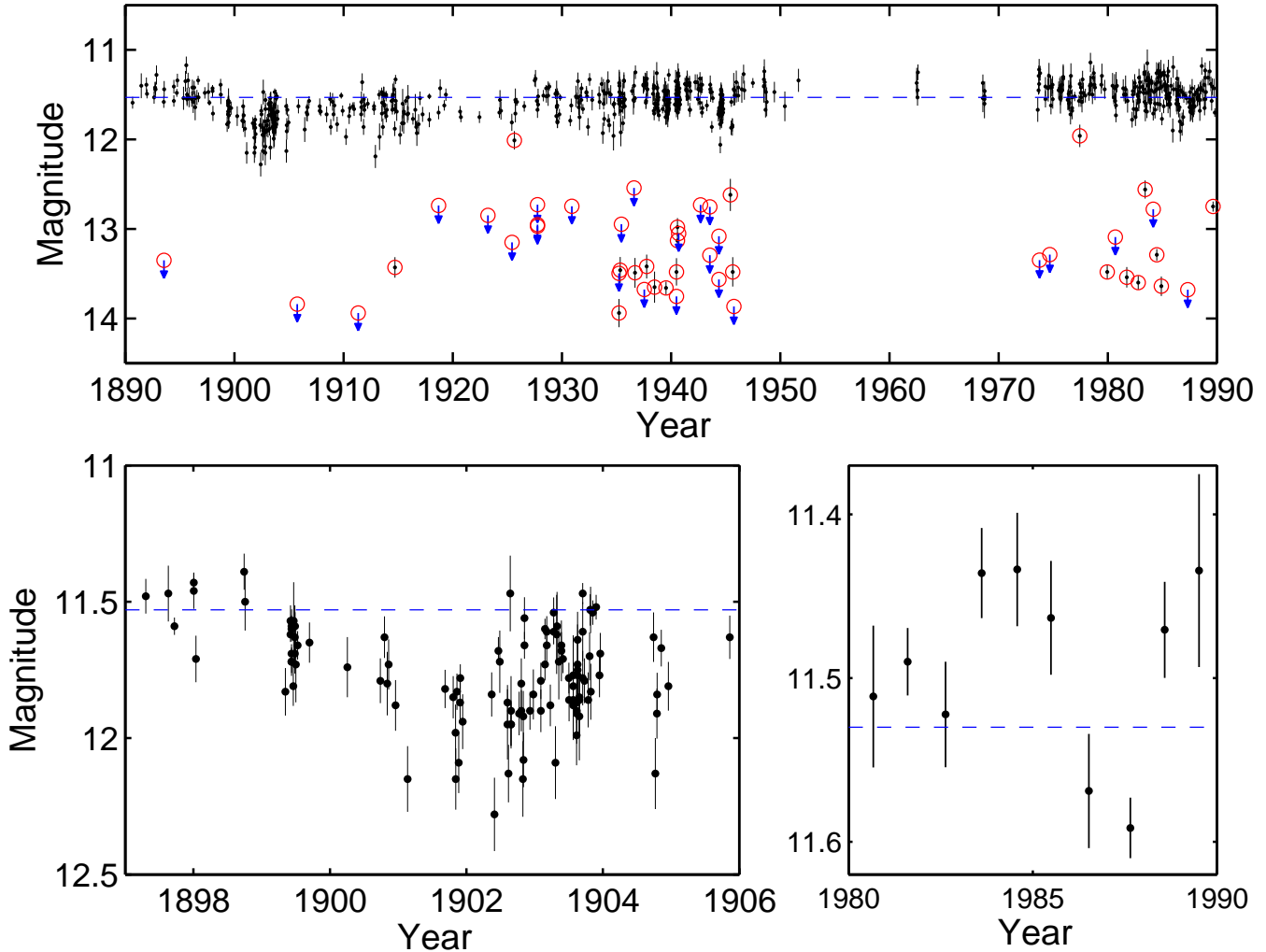, angle=0, width=\linewidth}
\caption{DASCH Light curve of KU Cyg. The upper panel shows the whole light curve from 1890 to 1990.
The lower-left panel shows the light curve from 1897 to 1906 for the dimming event,
and the lower-right panel shows the light curve from 1980 to 1990 binned in 1 year increments
with error bars showing the standard deviation of the mean (primary eclipses excluded).
Black dots with errorbars are DASCH measurements calibrated to g band, and blue arrows are upper limits.
Red circles mark primary eclipses with orbital phase $0.965-0.035$.
In all three panels, blue dashed lines show a constant magnitude of 11.53, which is the median magnitude
of KU Cyg from 1910 to 1990 with phases $0.05-0.95$ (excluding the primary eclipses).}
\end{figure*}

\begin{figure*}
\epsfig{file=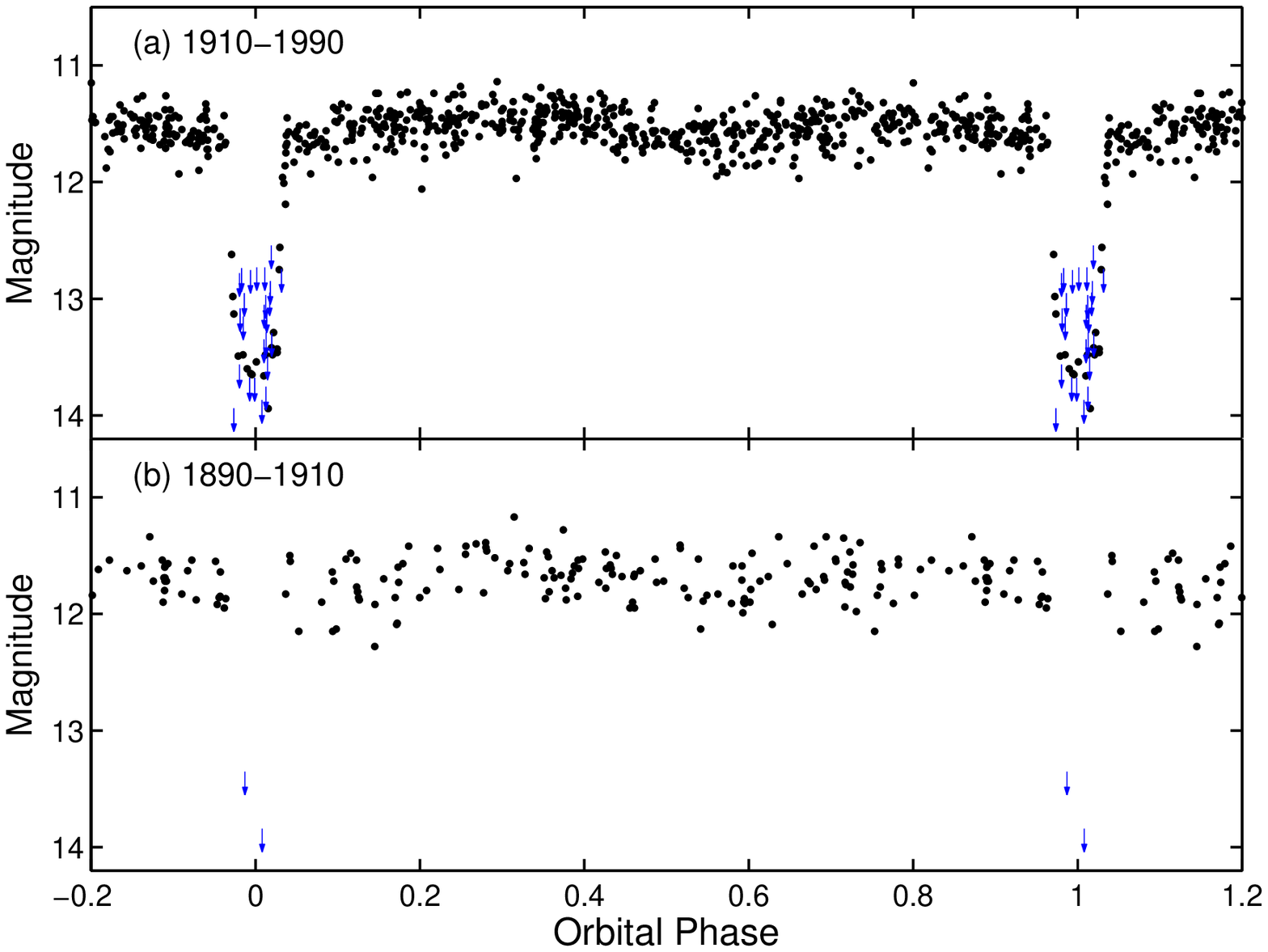, angle=0, width=\linewidth}
\caption{DASCH folded light curve of KU Cyg, using the adopted ephemeris from Olson (1988). The upper panel shows the folded light curve from 1910 to 1990,
and the lower panel shows the folded light curve from 1890 to 1910.
Black dots are DASCH measurements, and blue arrows are upper limits.}
\end{figure*}

\section{Discussion}
Since the orbital period of KU Cyg is $\sim38.4$ days, the $\sim5$-yr fading-brightening
event we found from 1899 to 1904 must come from sources other than eclipses.
Our plates are blue sensitive, where the flux is dominated by the F star;
This is also illustrated in the folded light curve, where there are up to $2.4$ mag decreases during the primary eclipses,
which means the K5III giant contributes $<10\%$ of the total flux.
Therefore, the $\sim0.5$ mag dimming must be related to the F star and its accretion disk.

\subsection{Higher mass transfer rate?}
The simplest explanation of the fading-brightening event, perhaps, is increasing disk extinction of the F star arising from increasing mass transfer rate.
RX Cas is an edge-on mass-transfering eclipsing binary system ($P=32.3$ days), similar to KU Cyg.
Kriz et al. (1980) suggested in RX Cas that
when the rate of mass transfer increases, the disk density and consequently optical depth increases,
leading to attenuation of the central star and resulting in fading in its light curve.
Similar effects have been observed in Be shell stars (edge-on Be stars with high $v\sin i$),
where fadings are seen in mass ejection events (see e.g. Huber \& Floquet 1998; Gutierrez-Soto et al. 2008).

It has been found that the accretion disk surrounding the F star in KU Cyg is thick and dusty,
with dust to gas ratio $\sim10^{-6}$, and a total disk mass $\sim10^{-8}-10^{-5}\ M_{\odot}$ (Olson 1988; Smak \& Plavec 1997).
Smak \& Plavec (1997) estimated an extinction of 1.3 mag in V of the F star by the disk,
with $\sim 0.75$ mag extinction due to the full obscuration of about half of the F star by the disk,
and the remaining $\sim 0.55$ mag due to the absorption in the disk atmosphere.
To cause a $0.5-0.6$ mag dimming observed in KU Cyg, and assuming the same gas-to-dust ratio,
the disk mass excess would be on the order of the accretion disk mass, i.e. $\sim10^{-8}-10^{-5}\ M_{\odot}$.
The timescale of accretion in the disk of KU Cyg is a few years (Smak \& Plavec 1997).
If we assume a $\sim10^{-8}-10^{-5}\ M_{\odot}$ ejection from the K5III star
arrives at the outer region of the accretion disk around 1899,
during the next few years, the ejected mass would move inwards, so that dust grains would be
levitated above the disk by radiation pressure and photophoresis (see e.g. Vinkovic 2009; Wurm \& Haack 2009).
This would result in an increased covering factor and a larger extinction.

Later, when the dust moves to smaller radii in the disk and gets closer to the F star,
it should reach the evaporation temperature and most dust particles at high scale heights would evaporate,
leading to a brightening.
Additionally, some of the dust particles may be transported outwards (Takeuchi \& Lin 2003; Ciesla 2007), cool down to condensate, and disperse,
leading to more extinction.
Moreover, when the clump of excess mass is accreted onto the F star, the energy release on the boundary layer also leads to a brightening,
as discussed in Olson (1988) and Smak \& Plavec (1997) to explain the $\sim0.2$ mag brightening bump around 1984.
The above three processes may together produce the fluctuations from 1903 to 1905 in KU Cyg.

\subsection{Higher dust to gas ratio?}
An alternative explanation is that the dimming is caused by dust excess rather than mass excess,
i.e., higher dust to gas ratio, in the disk surrounding the F star.
In a similar way to the above discussion, when the dust clumps move inwards and
are levitated above the disk,
extinction of the F star becomes larger, and the system becomes fainter.
The system brightens when the dust grains move closer to the F star and evaporate.

However, where the dust excess comes from is a challenging question.
It might arise from clumps of dusty gas ejected from the K5III giant.
We have discovered a group of K2III giants with $\sim1$ mag dimming over $10-100$ years,
which do not match any known classes of variables (Tang et al. 2010).
We suggested that the dimming might be caused by dust extinction in a certain evolution stage of K giants, though how the dust gets ejected is still unclear.
The dimming of KU Cyg in 1899 might arise from similar dust ejection processes in the K5III giant.
The dust clumps then move to the accretion disk around the F star through Roche Lobe overflow.

\subsection{Summary}
In summary, the 5-year dip in brightness of KU Cyg observed around 1900 is unique in the way that it faded slowly and brightened relatively quickly,
which is contrary to the dust production and dispersion process observed in RCB stars,
where it usually takes longer for the dust to disperse and thus longer for the star to brighten (Clayton 1996).
This fading-brightening event in KU Cyg is probably related to the accretion disk surrounding the F star,
which provides interesting clue for the study of dust confinement, levitation and evaporation in an accretion disk.

\acknowledgments
We are grateful to the anonymous referee for helpful comments and suggestions.
We thank Alison Doane, Jaime Pepper and Bob Simcoe at CfA for their work on DASCH,
and many volunteers who have helped digitize logbooks, clean and scan plates (\emph{http://hea-www.harvard.edu/DASCH/team.php}).
We are grateful to the logbook transcription team at the American Museum of Natural
History in New York City led by Michael Shara and Holly Barton.
S. T. thanks Ruth Murray-Clay, Matthew Holman, Philip Chang and Li Zeng for helpful discussions.
The DASCH project is supported by NSF grants AST-0407380 and AST-0909073. \\


\end{document}